\documentclass[]{aa} 
\usepackage{natbib}
\usepackage{graphicx}
\usepackage{color}
\usepackage{sidecap}
\usepackage{hyperref}
\begin{document}

\title{Variation of the Mn\,{\sc i}~539.4~nm line with the solar
cycle}
\titlerunning{The solar cycle variation of the Mn\,{\sc i}~539.4~nm line}
\author{S. Danilovic \inst{1}\and S.K. Solanki \inst{1,2}\and W. Livingston \inst{3}\and N. Krivova \inst{1}\and I. Vince \inst{4}}

\institute{ Max-Planck-Institut f\"ur Sonnensystemforschung,
Justus-von-Liebig-Weg 3, 37077 G\"ottingen, Germany \and School of
Space Research, Kyung Hee University, Yougin, Gyeonggi 446-701,
Korea \and
National Solar Observatory, 950 North Cherry Avenue, Tucson, AZ 85718, USA \and
Astronomical Observatory, Volgina 7, 11160 Belgrade 74, Serbia \\
\email{danilovic@mps.mpg.de}}

\date{Received / Accepted}
\abstract
{As a part of the long-term program at Kitt Peak National Observatory
(KPNO), the Mn\,{\sc i}~539.4~nm
line has been observed for nearly three solar cycles using the McMath
telescope and the 13.5 m spectrograph in double-pass mode. These
full-disk spectrophotometric observations revealed an unusually
strong change of this line's parameters over the solar cycle.}
{ Optical pumping by the Mg II k line was originally proposed to explain these variations. More recent studies have proposed that this is not required and that the magnetic variability (i.e., the changes in solar atmospheric structure due to faculae) might explain these changes. Magnetic variability is also the mechanism that drives the changes in total solar irradiance variations (TSI). With this work we investigate this proposition quantitatively by using the same model that was earlier successfully employed to reconstruct the irradiance.}
{We reconstructed the changes in the line parameters
using the model SATIRE-S, which takes only variations of the daily
surface distribution of the magnetic field into account. We applied
exactly the same model atmospheres and value of the free parameter as were used in previous solar irradiance reconstructions
to now model the variation in the Mn\,{\sc i}~539.4~nm line profile and in neighboring Fe\,{\sc i} lines. We compared the results of the theoretical model with KPNO
observations.}
{The changes in the Mn\,{\sc i}~539.4~nm line and a neighbouring Fe\,{\sc i}
539.52 nm line over approximately three solar cycles are reproduced
well by the model without additionally tweaking the model
parameters, if changes made to the instrument setup are taken into
account. The model slightly overestimates the change for the strong Fe\,{\sc i} 539.32 nm line.}
{Our result  confirms that optical pumping of the Mn\,{\sc ii}~539.4~nm line by
Mg II k is not the main cause of its solar cycle change. It also provides 
independent confirmation of solar irradiance models which are based on the assumption that irradiance variations are caused by
the evolution of the solar surface magnetic flux. The result obtained here also supports the spectral irradiance variations computed by these models.} 

\authorrunning{S. Danilovic et al.}
\titlerunning{The Solar Cycle Variation of the Mn\,{\sc i}~539.4~nm line}
\maketitle

\section{Introduction}

Solar irradiance varies on timescales of
minutes to decades. For variations on timescales extending from a day to the length
of the solar cycle, different causes have been proposed: changes of
the internal thermal structure of the Sun, changes of subsurface fields, or
changes of the field at the surface. The third explanation has found
considerable support \citep{Solanki:etal:2005,Domingo:etal:2009,Solanki:etal:2013}. 

In particular, the Spectral And Total Irradiance REconstruction for the satellite era (SATIRE-S) has been
successfully used to describe the variability in the changes
of total irradiance during the past three cycles. At least $92\%$ of the
total solar irradiance (TSI) variability can be assigned to
changes in the magnetic field distribution at the solar surface
\citep{Krivova:etal:2003,Wenzler:etal:2005,Wenzler:etal:2006,Krivova:etal:2011,Ball:etal:2012,Kok:etal:2014}.
However, since in its present form the SATIRE-S model contains a free
parameter, these conclusions may not be entirely compelling. It is therefore important
to reproduce spectral features as well. This has been done for the
Variability of IRradiance \& Gravity
Oscillations \citep[VIRGO;][]{Froehlich:etal:1995} color channels by
\citet{Krivova:etal:2003}, UV irradiance \citep{Krivova:etal:2006,Krivova:etal:2009,Kok:etal:2014},
and for spectral irradiance over a broad wavelength range on solar
rotation timescales \citep{Unruh:etal:2008}. 

The recent spectral data from 
the Spectral Irradiance Monitor (SIM) on SORCE over the declining phase of cycle 
23 \citep[cf.][]{Harder:etal:2009}, however, showed solar cycle trends that are significantly 
different from those expected both from earlier measurements and from models \citep{ermolli,Ball:etal:2012,Kok:etal:2014}. In particular, irradiance in the visible between 400 and 700 nm 
was found to increase with the decrease of the TSI. In this paper, we
use the SATIRE-S model to describe the solar cycle variability of Mn\,{\sc i}
539.47 nm and two neighbouring neutral iron lines, which sets
additional constraints on the model. Modeling the solar cycle
variation of spectral lines is of particular importance since
spectral lines are estimated to contribute more than $50-90\%$ of the TSI
variations over the solar cycle
\citep{Mitchell:Livingston:1991,Unruh:etal:1999}. Moreover, some studies have even suggested that the solar irradiance variability in the UV, violet, blue, and green spectral
domains is fully controlled by the Fraunhofer lines \citep{Shapiro:etal:2015}. 

The Mn\,{\sc i}~539.47~nm line is interesting for several reasons. The
hyperfine broadening, due to the interaction of the electronic
shell with the nuclear spin, makes it relatively insensitive to nonthermal
motions in the photosphere \citep{Elste:Teske:1978,Elste:1986}.
This was recently demonstrated by \citet{Vitas:etal:2009} with 3D magnetohydrodynamic (MHD) simulations. They showed that the Mn\,{\sc i}~539.47~nm line is not
smeared by convective motions, in contrast to the neighboring narrow Fe\,{\sc i} 539.52 nm line, which  always becomes weaker as a result of the smearing. The Mn\,{\sc i} line, on the other hand, becomes stronger in
sunspots and weaker in plages and the network, as observed
by \cite{Vince:etal:2005a,Vince:etal:2005b} and \cite{Malanushenko:etal:2004}.
This means that although the Mn\,{\sc i} line is formed in the photosphere
\citep{Gurtovenko:Kostyk:1989,Vitas:2005}, the 'Sun-as-a-star'
observations \citep{Livingston:Wallace:1987,Livingston:1992} of this\,{\sc } line 
exhibit a significant cycle dependence that is not typical for other photospheric lines recorded in parallel. A time-series
analysis shows that even variations
on the solar rotation timescale can be found around the solar activity
maximum \citep{Danilovic:etal:2005}. One explanation of this chromospheric-like behavior was
earlier proposed by \citet{Doyle:etal:2001}. They found that optical
pumping by Mg\,{\sc ii}~k might be the reason for this Mn line to mimic
the change seen in Mg\,{\sc ii}~k. However, the analysis of
\citet{Vitas:Vince:2007} proved that the Mn\,{\sc i}~539.47~nm line is
insensitive to the photons emitted in the cores of Mg\,{\sc ii}~h and k.

\citet{da:vi:2005} showed that extended the modeling of the 'Sun-as-a-Star' spectra of Mn\,{\sc i}  539.47 nm could explain the change observed over a short period of solar cycle 23. Later, \citet{Vitas:etal:2009} demonstrated the change in the Mn\,{\sc i} line in magnetic flux concentrations in one MHD snapshot, but they did not model the 'Sun-as-a-Star' spectra. In this paper we model the Mn\,{\sc i} cycle change over the full period of 1978-2009 that was covered by Livingston's 'Sun-as-a-Star' monitoring program\footnote{\url{ftp://vso.nso.edu/cycle_spectra/reduced_data/dat2.5394}} \citep{Livingston:etal:2010}. In addition, we  analyze the behavior of the two neighboring Fe\,{\sc i} lines that were
observed in the same spectral channel.

The structure of the paper is as follows: observations and modeling technique are described in Sect.~\ref{sec:obs}.
The disk-averaged line synthesis is treated in
Sect.~\ref{sec:lines}. The calculated center-to-limb variations of
the line profiles are given in Sect.~\ref{sec:profiles}. Results
for different observational phases (before and after 1992) and for three solar cycles are discussed in
Sect.~\ref{sec:timesets}. Concluding remarks are given in
Sect.~\ref{sec:conclusions}.

\section{Observational data and modeling technique}
\label{sec:obs}

\subsection{Data}
\label{data}

The observational data modeled here were obtained with the 13.5 m scanning spectrometer set in double-pass
mode and mounted on the Kitt
Peak McMath telescope \citep{Brault:etal:1971}. We describe the observations in some detail since no comprehensive description exists in the
literature.

The sunlight is delivered to the
spectrograph without any prefocusing. There a 0.5x10 mm entrance
slit forms a pinhole image of the Sun on the grating (camera
obscura principle). This specific instrumental set produced the disk-integrated or 'Sun-as-a-star' spectrum.

The $2$~\AA\  wide spectral range around the Mn\,{\sc i}~539.47~nm line was
observed in fifth order of the grating, which yielded a spectral
resolution of around 106000. Line intensities and equivalent
widths were automatically extracted using the data reduction
program developed by J. Brault at Kitt Peak
\citep{Brault:etal:1971}. For the continuum normalization the data
point at 539.491 nm was used because the comparison with the Kitt
Peak spectral atlas of \citet{Wallace:etal:2007} showed that it is
entirely free of telluric blends, whereas most other
(pseudo-)continuum wavelengths are affected. The intensity values
at this wavelength were normalized to unity. This might be slightly
too high, as a comparison with spectra acquired with the Fourier Transform Spectrometer (FTS) reveals
(Fig.~\ref{fig:profil}). However, tests have shown that any small
error introduced by this normalization does not affect the results
in a significant way. A quadratic fit to the cores of each of
the lines was made, and the line central depth was defined as the
difference between the minimum of the fit and the normalized continuum
value. The equivalent width was calculated within given wavelength
ranges (shown below in Fig.~\ref{fig:profil}).

A first set of observations with an unchanged setup was taken from
1979 to 1992, with recordings being made a few times each month. In
1992 a new larger grating with dimensions $42\times32$~cm$^{2}$
was installed, compared with $25\times15$~cm$^{2}$ for the grating
mounted before 1992. This instrumental change had a measurable
effect on the data. Before the change, part of the solar limb fell
outside the grating, whereas with the new grating, the full solar
disk was sampled. This introduced a jump in the measured
equivalent width (EW) and central depth (CD) of lines obtained
before and after 1992. After mounting, an experimental period with the
new grating lasted until 1996, when the system was fixed and
observations were performed regularly again. However, constant
realignment of the system between 1996 and 1998 led to an
increased variance in the parameters of the recorded spectral
lines. As a result, we analyzed and modeled observations recorded
on 465 days from January 1979 to September 1992 and on 463 days
in the period from September 1998 to October 2009. Between September
1992 and September 1998, observations were obtained on 45 days.
These were not taken into account in the work described here because of the varying settings.

\subsection{SATIRE-S model}
\label{model}

The SATIRE model
described by  \citet{Fligge:etal:2000} and
\citet{Krivova:etal:2003} assumes that the change in solar irradiance can
be explained by the evolution of the solar surface magnetic field. The
model categorizes all solar surface features into four groups based on
their brightness and magnetic flux density. The surface area with
a brightness below a given threshold is identified as the area covered
by sunspots. These are further partitioned into umbrae and
penumbrae, depending on the brightness level. Facular regions are
classified as areas that have a magnetic flux density above a
threshold, but do not belong to sunspots or pores. The areas that do not
fulfil any of these criteria are classified as quiet-Sun areas. To identify the
features in the period from 1978 to 2003, we used full-disk
magnetograms and continuum images recorded with the 512 channel
magnetograph and the spectromagnetograph, mounted on the Kitt Peak
Vacuum Tower Telescope. Magnetograms obtained with MDI/SOHO were used for the later period. A detailed description of data
calibration is given in
\citet{Wenzler:etal:2004,Wenzler:etal:2006} and \cite{Ball:etal:2012}.

Since the spatial resolution of the maps is on the order of a few
arcsec, the model takes into account that magnetic
elements in faculae are not fully resolved; to do this, a
filling factor is introduced. It increases linearly with magnetic flux density
and reaches unity at a magnetogram signal of $B_{\rm sat}$. For
$B>B_{\rm sat}$ the filling factor remains constant at 1. $B_{\rm sat}$ is
the only free parameter in the SATIRE-S model. The value of
$B_{\rm sat}~=~330~G$ used here is the same as in
\citet{Ball:etal:2012}. 

The stratification of physical parameters with depth was assumed to
remain unchanged over the solar cycle and was represented by 1D model atmospheres \citep{Unruh:etal:1999}. The Kurucz
standard solar atmosphere \citep{Kurucz:1991,Kurucz:1992} was used for the quiet Sun, and stellar
models with a gravitational acceleration of log~g$=4.5$~m/s$^{2}$
and effective temperature of $4500$~K and $5400$~K were used for
umbrae and penumbrae, respectively. As the model atmosphere for the
faculae, \cite{Unruh:etal:1999}adopted FALP from \citet{Fontenla:etal:1999}, but without
the chromospheric temperature increase and with a slightly smaller
temperature gradient in the deeper atmospheric layers. As demonstrated in that paper, these
changes provide a good agreement of synthesized and observed
facular contrast in the optical range.

\begin{table*}
\caption{Line characteristics: wavelength, transition, oscillator
strength, and excitation energy.} \label{tab:char} \centering
\begin{tabular}{l c c c c}
\hline \hline
line & transition & log $gf$ & $\chi ~[eV]$ \\
\hline
Fe\,{\sc i} 539.3167 nm & $z^{5}D_{3}-e^{5}D_{4}$ & $-0.762$ & $3.24$\\
Mn\,{\sc i} 539.4677 nm & $a^{6}S_{5/2}-z^{8}P^{0}_{7/2}$ &  $-3.453$ & $0.0$\\
Fe\,{\sc i} 539.5215 nm & $z^{5}G_{2}-g^{5}F_{1}$ & $-1.763$ & $4.44$\\
\hline
\end{tabular}
\end{table*}

\begin{figure}
    \centering
    \includegraphics[width=0.7\hsize]{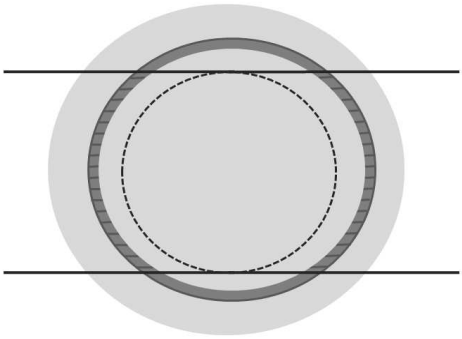}
    \caption{Illustration of flux calculations for the period before 1993 when the old grating was used. The two horizontal lines delimit the part of the disk that illuminates the grating.
    The dashed circle corresponds to $\mu=\mu_{\rm cut}$.}
    \label{fig:schema}
\end{figure}

\subsection{Spectral line calculations}
\label{sec:lines}

The KP spectra containing the Mn\,{\sc i}~539.47~nm line also include two
Fe\,{\sc i} lines. These lines exhibit smaller variations over the solar
cycle, allowing them to be partially used to check for changes in
instrumentation etc. The main characteristics of all three lines,
taken from the VALD database \citep{Kupka:etal:1999}, are given in
Table~\ref{tab:char}.

The procedure of calculating disk-integrated flux in this spectral
range is as follows: Time-independent emergent intensities of the
Mn\,{\sc i} line and the neighboring Fe\,{\sc i} lines are calculated  in local thermal equilibrium (LTE)
for each model atmosphere for various heliocentric angles  using
the SPINOR code \citep{Frutiger:etal:2000}. No magnetic field is
introduced in the calculations. The hyperfine structure of the Mn
I line is included as blends with displacements calculated using
the hyperfine constants from \citet{Davis:etal:1971} and
\citet{Brodzinski:etal:1987} and relative intensities of the
components from \citet{Condon:Shortley:1963}. Components that are
closely spaced with respect to the total splitting were
combined to reduce the computing effort. Emergent intensities,
output from SPINOR, were then combined with the relative
contributions of different features for different heliocentric
angles.

Before 1993, the incomplete sampling of the solar disk by the
grating had to be taken into account. Figure~\ref{fig:schema}
illustrates the simple method that we applied. We assumed that the
part of the image of the solar disk bounded by the two horizontal
lines covers the grating, while the rest is lost due to the size of
the grating. Let $\mu_{\rm cut}$ be the cosine of the a priori unknown
heliocentric angle for which the disk image fully covers the
grating. Flux coming from the region shaded dark
gray in the figure, for example, which corresponds to $\mu<\mu_{\rm cut}$, is
partially lost. For each of these 2-degree-wide regions we
introduced coefficients proportional to the annular surfaces that
lie in between the horizontal lines (shaded with short horizontal
lines). The radiative flux coming from the whole annulus is
multiplied by this coefficient. The coefficients are normalized
such that the sum over all coefficients for every region at
$\mu<\mu_{\rm cut}$ is unity. In this way, we did not specify which
parts of the solar disk (heliographic latitudes and
longitudes) were sampled and which were not. This treatment is
appropriate since the grating received light from different parts of the
solar disk during the day because the McMath telescope is fed
by a heliostat. Finally, the total flux at each
wavelength position was obtained by summing the fluxes
multiplied by the corresponding coefficients from all heliocentric
angles.

\begin{figure*}
    \centering
    \includegraphics[width=0.8\hsize, trim=0cm 0cm 0cm 0cm,clip=true]{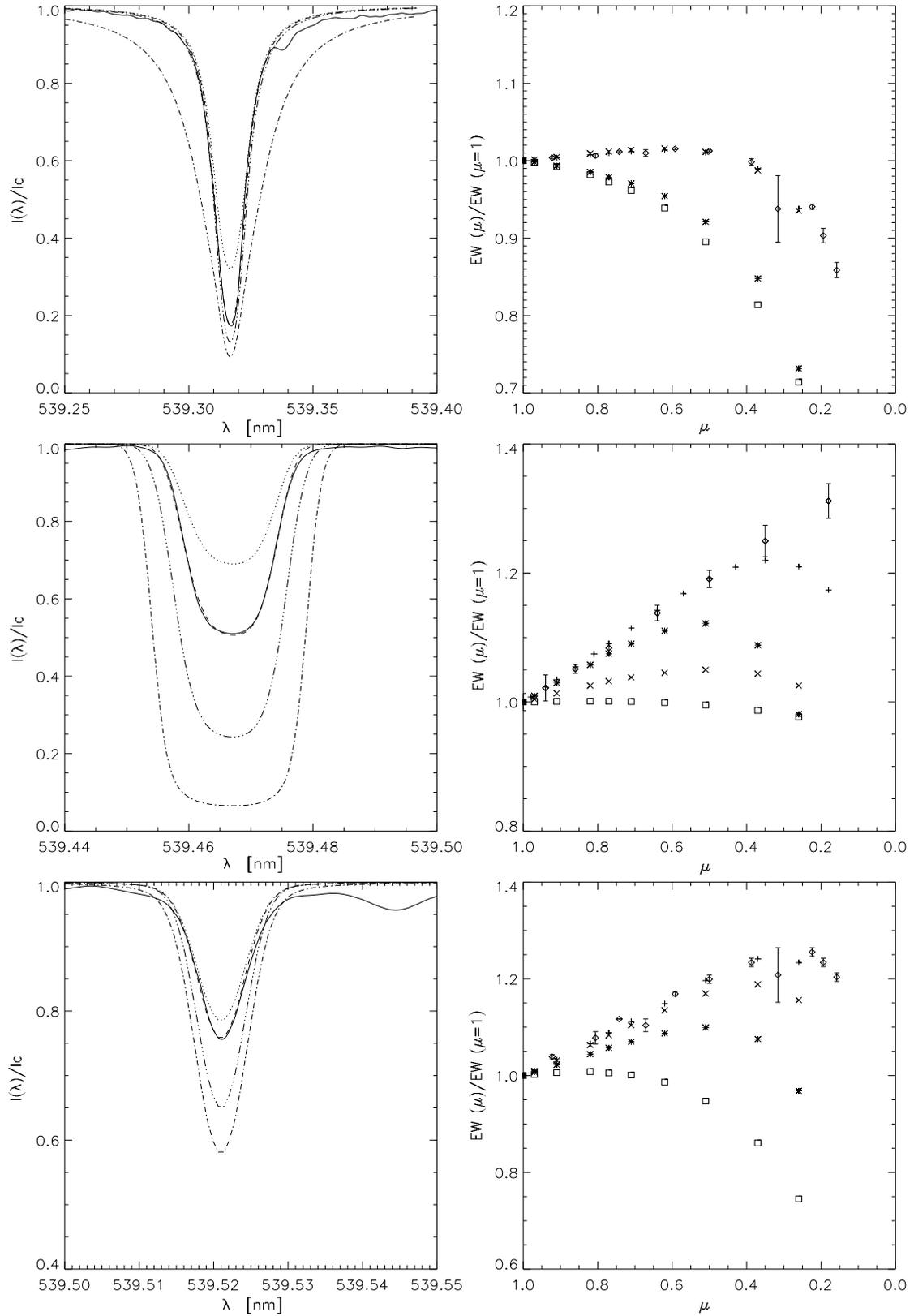}
    
    \caption{\textit{Left-hand side}: Profiles of Fe\,{\sc i} 539.32 nm,
    Mn\,{\sc i}~539.47~nm, and Fe\,{\sc i} 539.52 nm lines at the disk center, from top to bottom.
    The profiles are computed in the following model atmospheres: quiet Sun (dashed), faculae (dotted), penumbrae (double-dot-dashed), and umbrae (dash-dotted).
    FTS atlas profiles are overplotted (solid curves).
    \textit{Right-hand side}: Center-to-limb variation of the
EW of the lines.
    Symbols distinguish between the EW resulting from different model atmospheres: quiet Sun (plus signs), faculae (asterisks), penumbrae (crosses), and umbrae (squares).
    Diamonds with error bars are values observed in the quiet Sun.}
    \label{fig:cl}
\end{figure*}

\section{Results}
\label{sec:results}

Before computing the time series of the CD and the EW of the disk-integrated profiles of the two Fe\,{\sc i} lines and the
Mn line, we first verified that the lines were computed
properly, that is, that the measured profiles of line intensity were
accurately reproduced. We first computed the lines at disk center
and determined their contribution functions and log $gf\epsilon$ values
($gf$ is the oscillator strength times the statistical weight of
the lower level of the transition, and $\epsilon$ is the
abundance). In the second step we compared their center-to-limb
change (CLV) with measurements, and finally, we computed the time
series of the flux computed with the SATIRE-S model.

\subsection{{\bf Line profiles at disk center}}
\label{sec:profiles}

Employing elemental abundances from the literature and a quiet-Sun model atmosphere, we reproduced the FTS atlas profiles
\citep{Wallace:etal:2007} that were recorded in the quiet Sun at the disk
center by allowing the oscillator strengths and macroturbulence
broadening velocity to vary. A Gaussian profile was chosen for the
macroturbulence. The best-fit values are given in
Table~\ref{tab:dc}. The chosen abundances of Mn and Fe, which were
kept constant, are $\log\epsilon_{\rm Mn}=3.39$ and
$\log\epsilon_{\rm Fe}=7.5$
\citep{Bergemann:Gehren:2007,Shchukina:Bueno:2001}.  For the strong Fe\,{\sc i} 539.32~nm line, we introduced a damping
enhancement factor of 3 to fit the line wings. The fitted $\log
gf$ values differ from those taken from the VALD
(Table~\ref{tab:char}) by about 0.1 dex or less. The computed and
measured line profiles agree rather well (except in the wings,
where these lines are influenced by blends), as Fig. ~\ref{fig:cl} shows (panels on the left-hand-side; compare the solid with the dashed
line). The best-fit oscillator strengths and macroturbulence
values were retained when line profiles were subsequently calculated
in all other model atmospheres. The results are shown in the same
figure. All three lines display qualitatively the same dependence
on temperature. They become stronger in umbrae and penumbrae and
weaker in faculae. The magnitude of the temperature dependence is
quite different, however, with the Mn line showing by far the
highest temperature sensitivity, both in CD and in EW. Resulting formation heights (fh) of the line cores in the
quiet-Sun model atmosphere, as deduced from line depression
contribution functions
\citep{Magain:1986,GrossmannDoerth:etal:1988}, are given in
Table~\ref{tab:dc}. The strong Fe\,{\sc i}~539.32~nm line is thus formed
significantly higher than its neighbors. The obtained heights of
formation are in agreement with the previously determined values
\citep{Balthasar:1988,Gurtovenko:Kostyk:1989,Vitas:2005}.

\subsection{{\bf Center-to-limb variation}}
\label{sec:cl}

In a next step we computed the center-to-limb behavior of the
three lines and compared it with center-to-limb observations. For this we used the only two data sets available for these lines that
were reported by \citet{Hidalgo:etal:1994} and \citet{Balthasar:1988} for the Mn and Fe
lines, respectively. The calculated and observed EW
changes from disk center to the limb are shown in Fig.
~\ref{fig:cl} (right-hand panels).

The variations in the center-to-limb behavior of the lines reflect
their different sensitivities to the physical parameters and
velocity fields. The latter is the important difference between Fe
and Mn lines. As shown by \citet{Asplund:etal:2000}, to
reproduce the observed Fe line profiles at various heliocentric
angles, a more realistic
representation of the solar atmosphere has to be taken into account.
This would need to include surface
convection and hence would properly reproduce the line broadening due to small- and
large-scale velocity fields. Since we used 1D
plane-parallel models, we followed the classical method and
introduced a microturbulence that linearly increases with heliocentric
angle. In this way, the larger nonthermal broadening toward the
limb is mimicked, which comes from the high horizontal velocities in the
granulation. The values that best match the
observations are of the same order as found by
\citet{Holweger:etal:1978}. For the strong Fe line, the microturbulence increases
from $0.1$ km/s to $1$ km/s, while for the weak Fe\,{\sc i}~539.52~nm
line it grows from $1.5$ km/s to $2.4$ km/s (we assumed a linear
increase with $\mu$). These values were determined for the quiet Sun
and then maintained for the other model atmospheres.

The Mn line, in contrast, is intrinsically broad and hence all
broadenings caused by velocity fields are negligible \citep{Elste:1986,Vitas:etal:2009}. The
synthesized CLV for this line in the quiet-Sun
model atmosphere agrees with observations without any
additional change in microturbulence.
\begin{table}
\caption{Fitted values of oscillator strengths and macroturbulence
for the disk center. The rightmost column lists the formation
height of the line cores in the quiet-Sun atmosphere.}
\label{tab:dc} \centering
\begin{tabular}{l c c c c}
\hline \hline
line & log $gf$ (fitted) & $v_{\rm macro}$ [km/s] & fh [km]\\
\hline
Fe\,{\sc i} 539.3167 nm &  $-0.723$ & $1.389$ & $397$ \\
Mn\,{\sc i} 539.4677 nm &  $-3.471$ & $0.922$ & $197$ \\
Fe\,{\sc i} 539.5215 nm &  $-1.649$ & $0.916$ & $111$ \\
\hline
\end{tabular}
\end{table}

\begin{table}
\caption{Combined macroturbulence and instrumental broadening
velocity needed to reproduce the line profiles in periods before
and after the change of grating, respectively.} \label{tab:macro}
\centering
\begin{tabular}{l c c c}
\hline \hline
line & before [km/s] & after [km/s] \\
\hline
Fe\,{\sc i}~539.32~nm & 2.04 & 1.69  \\
Mn\,{\sc i}~539.47~nm & 2.86 & 2.68 \\
Fe\,{\sc i}~539.52~nm & 2.04 & 1.87  \\
\hline
\end{tabular}
\end{table}

\begin{figure}
    \centering
    \includegraphics[width=0.85\hsize]{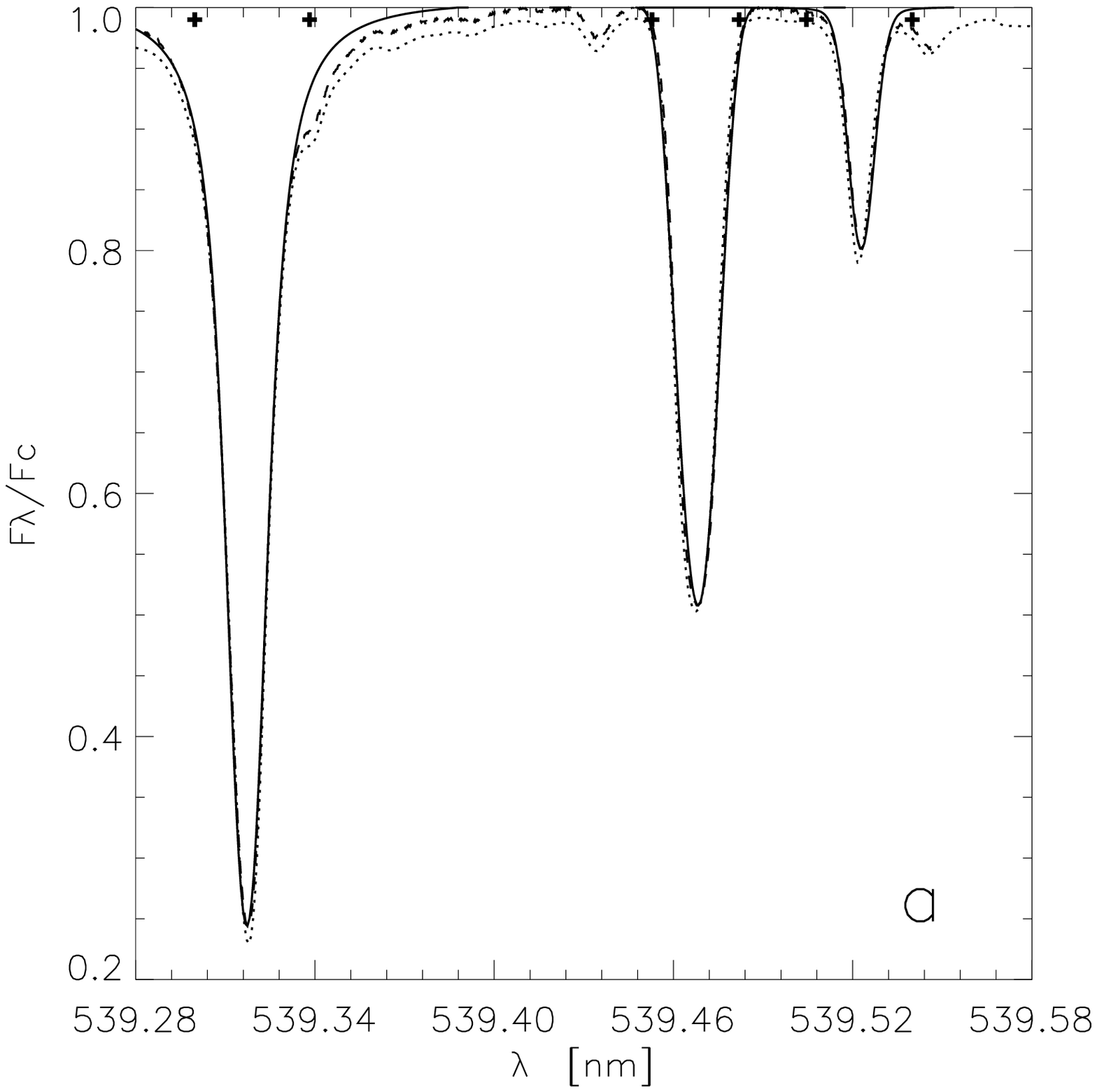}
    \includegraphics[width=0.85\hsize]{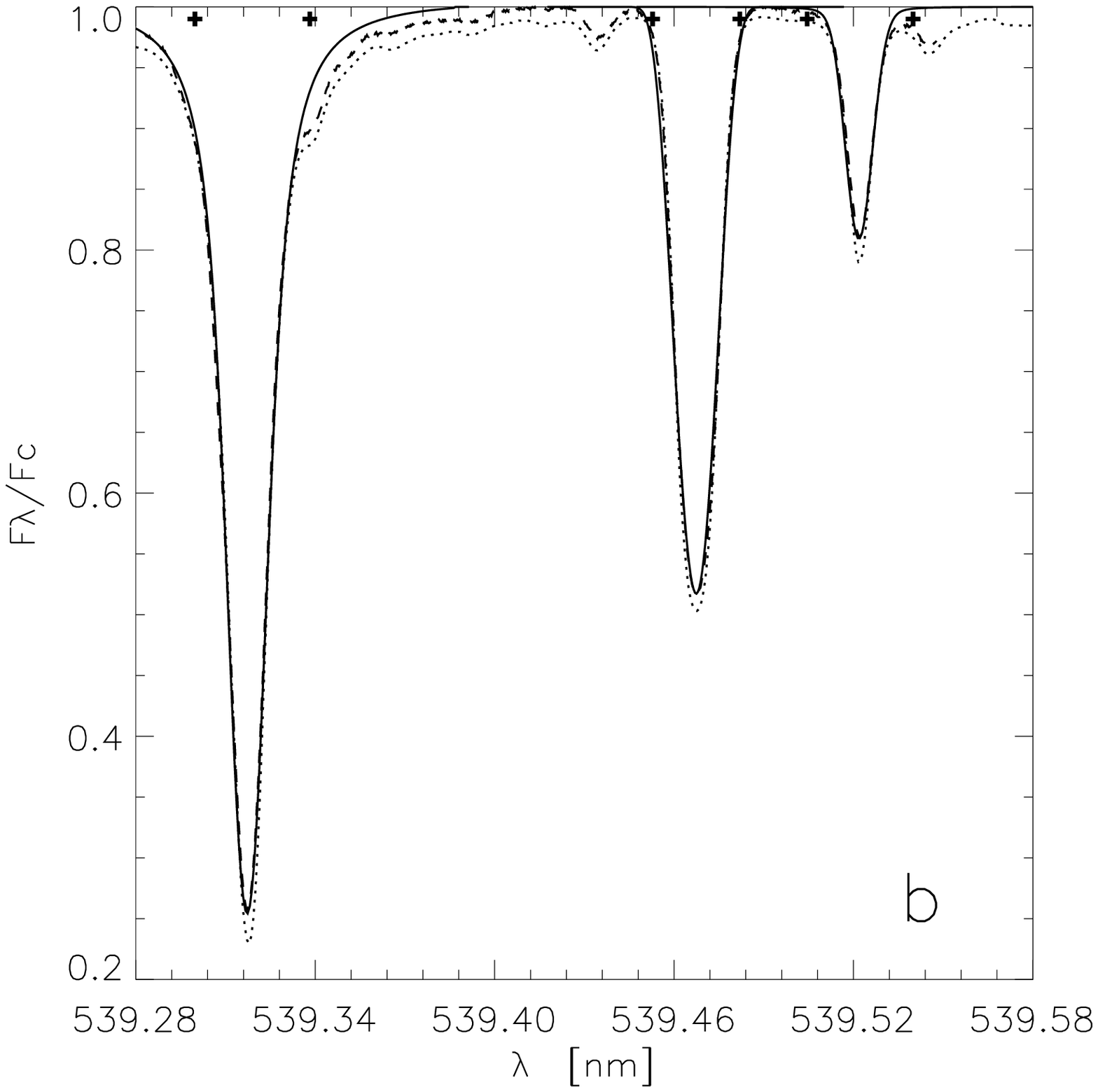}
    \caption{Synthesized line profiles (solid) and the same spectral range observed (dashed) on May 19, 1999 (upper panel) and May 8, 1986 (lower panel).
    The disk-averaged FTS atlas is overplotted in both panels (dotted lines).
    Crosses mark the wavelength ranges used for calculating the line equivalent widths.}
    \label{fig:profil}
\end{figure}

\begin{figure*}
    \centering
    \includegraphics[width=0.8\linewidth,trim= 0cm 0cm 0cm 0cm,clip=true]{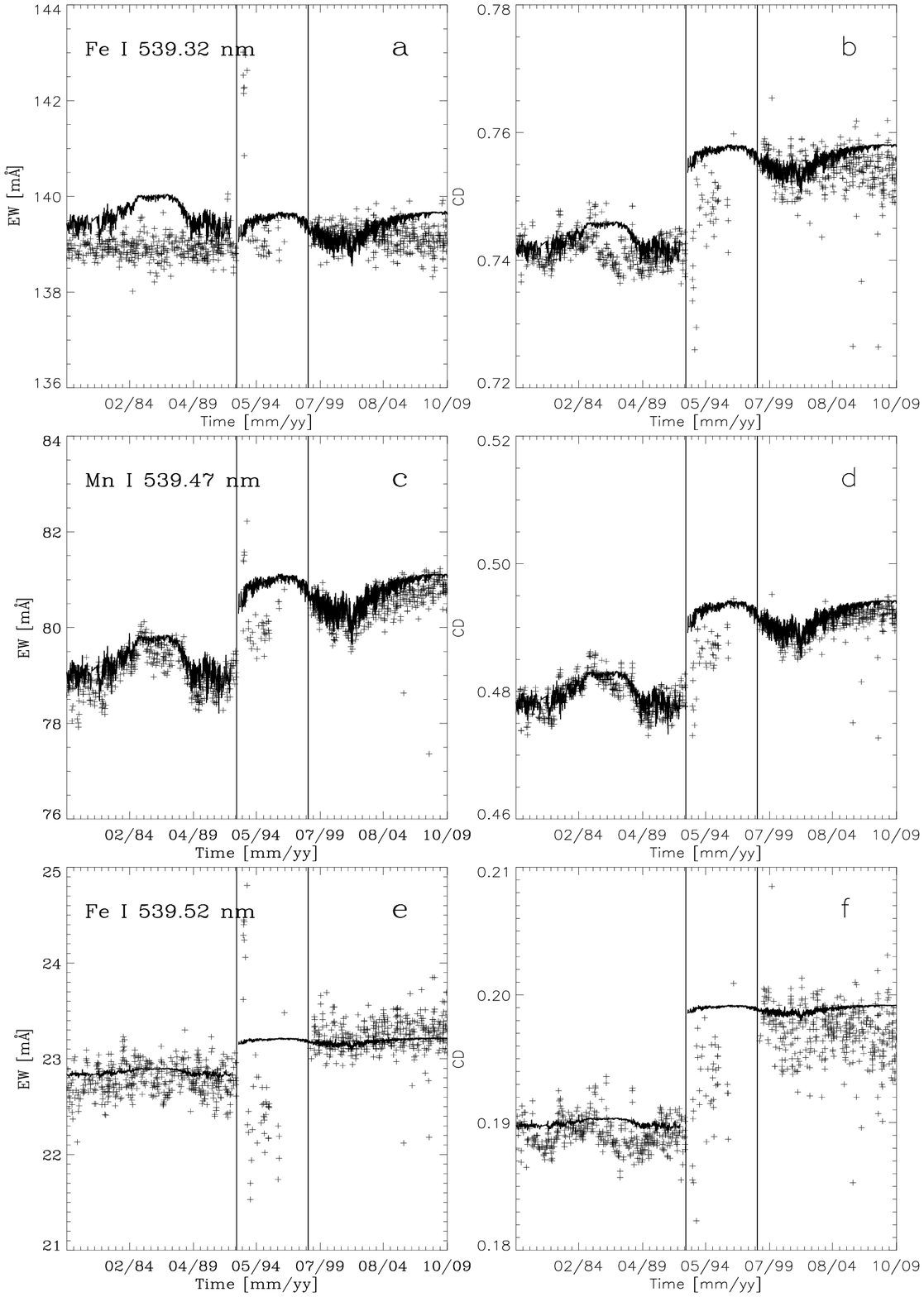}
    \vskip 2 pt
    \caption{Equivalent widths (left) and central depths (right)
    of the Fe\,{\sc i}~539.32~nm (top), Mn\,{\sc i}~539.47~nm (middle), and Fe\,{\sc i}~539.52~nm line (bottom) extracted from KPNO observations (crosses) and calculated using our model (solid
line). The experimental period (with changing setup parameters) is delimited by vertical lines. The modeled
EW values of the Fe\,{\sc i}~539.52~nm line are shifted for 0.7 m\AA.}
    \label{fig:time}
\end{figure*}

\subsection{Time-series comparison}
\label{sec:timesets}

Figure \ref{fig:profil} shows the whole observed spectral range,
obtained during two representative quiet days in the period
before and after the change of grating. The disk-averaged FTS
atlas \citep{Neckel:1999}  is overplotted, so that the change of
the line profiles that is due to the grating change is evident. The
central depths of all three lines were noticeably smaller before
the grating change. The remaining difference between the atlas and
observations taken after the grating change is due to the
continuum normalization. Crosses mark the spectral ranges used to calculate the
EW. They are chosen such that the calculated EW match
the EW observed in periods of quiet Sun after 1998. They are kept
unchanged for the whole modeled time interval.

Observations taken during the days with the lowest solar activity,
in the period before the change of grating (May 1986), were used to find
$\mu_{\rm cut}$. The comparison of the synthesized and observed
EWs gave a value of 0.7 for $\mu_{\rm cut}$. For the period after the change of the grating we set $\mu_{\rm cut}=0$.

All synthesized profiles were then convolved with Gaussians so that
the instrumental broadening, broadening due to solar rotation, and
convective velocity fields were taken into account. These effects
are thus approximated by a macroturbulence velocity, in addition
to the microturbulence, which was employed unchanged from the
values deduced in Sect.~\ref{sec:cl}. The broadenings applied to
the line profiles were adjusted for each line separately, so
that we reproduced the observed line central depths taken at low solar
activity. The resulting Gaussian half-widths are given in Table
\ref{tab:macro}. The line broadening was determined separately
for the periods before and after the grating change, with values
for the latter period being lower. Several factors can play a
role here. First, the change of grating and set-up produces a
change in the instrumental broadening. Second, by excluding the
outer parts of the disk, the influence of the parts of the solar
disk that produce more shifted line profiles as a result of solar rotation
is changed. Third, before the grating change fewer
regions close to the limb were sampled, which experience larger broadening due
to convection. As a result of the superposition of these effects,
the lines show a $20\%$ larger broadening on average before 1993
(implying a broadening velocity higher by about $10\%$ ). Finally,
errors introduced in the disk-averaged profiles as a result of the
simplicity of our modeling cannot be ruled out, but cannot be
estimated either. Assuming LTE also introduces errors.

Figure \ref{fig:time} shows the observed and modeled change of
the EW and CD of the three lines over nearly three solar cycles. Vertical lines
enclose the period during which the setup of the telescope was
frequently changed. The modeled values of the Fe\,{\sc i} 539.52 nm line
EWs were systematically lower because of the slight
blend in its vicinity (see Fig. \ref{fig:profil}), therefore we
corrected for them by adding a constant of $0.7$ m\AA.

It is most conspicuous in Fig.~\ref{fig:time} that
the shift introduced by the change in the grating is relatively
well modeled by the simple technique that we applied. With a
single free parameter $\mu_{\rm cut}$, we were able to produce the
offset in three parameters, the EWs of the three lines.

Another feature in Fig.~\ref{fig:time} is the observed
variability of the Mn\,{\sc i} line is stronger than in the Fe\,{\sc i} lines. The variability
of the Mn\,{\sc i} 539.47~nm line is well reproduced, showing the correct
magnitude of the dips of both EW and CD during periods of high
activity. The modeled EW and CD of the weak Fe\,{\sc i} 539.52~nm line
show only a weak change over the solar cycle and match the
observations relatively well. The modeled variations are smaller
than the scatter of the data. The strong Fe\,{\sc i} line, on the other
hand, shows a significantly stronger modeled change than is observed.
This could be a consequence of the LTE assumption that we
made. As a result of the greater formation height of this line (see Table
\ref{tab:dc}), it samples a considerably larger temperature
difference between the quiet Sun and the facular model atmosphere
(of around 500~K). In LTE this largely compensates for the
relative temperature insensitivity of this line. The overestimate
of the Fe~I~539.32~nm line variability suggests that one of the
assumptions made by the model is not met. Given the strength of
the line and its high formation height, the departures from LTE
might be the main cause.

In total, observations and simulations overlap for
184 and 456 days before and after the change of
grating, respectively. Correlation coefficients between observed and computed EW of the Mn line are 0.86 and 0.83, while for the CD  they are 0.81 and 0.82 for two periods, respectively. The modeled EW better
correlates with observations taken before 1992 than after. One
reason is that the observational values obtained after 1993 have
intrinsically larger scattering \citep{Livingston:etal:2007}. 

 To some extent as a result of the lack of clear variability in the observations,  there is evidently no correlation between the observed and modeled data of the Fe lines, therefore we omit reporting the corresponding correlation coefficients. 

\section{Conclusions}        %
\label{sec:conclusions}     %

We calculated the change of the disk-integrated Mn\,{\sc i}~539.47~nm line and
two neighboring Fe~I lines from 1979 to 2009 by using the SATIRE-S
model. This model has so far reproduced variations of total and
spectral irradiance on timescales of days to multiple solar
cycles. However, it has never been tested on individual spectral
lines. This test is of particular interest (1) because of the
evidence that spectral lines are the dominant contributors to
TSI variations over the solar cycle
\citep{Mitchell:Livingston:1991,Unruh:etal:1999,Shapiro:etal:2015} and (2) because the recent 
data from SORCE/SIM \citep{Harder:etal:2009} suggest that the irradiance 
in the visible varies in antiphase with changes in the TSI. 

It was noted previously \citep{Krivova:etal:2006,Unruh:etal:2008} that
SATIRE-S might overestimate the variations over the solar
cycle of stronger lines that are formed
higher up, possibly because NLTE effects are neglected.
Indeed, the variation in the profile of the strong Fe~I~539.32~nm line modeled here, in particular in the EW, is also overestimated.

At the same time, the observed scatter for the weak Fe~I~539.52~nm line is much larger than the modeled change in the line parameters. This is similar to the result found by \cite{Penza:etal:2006}, who analyzed the second set of photospheric lines observed as part of the 'Sun-as-a-Star' program. They employed a simpler model that neglected the heliocentric dependence of the emergent spectra and took certain proxies of solar magnetic activity to estimate the coverage of the solar disk by different features, whereas we employed direct high-resolution full-disk observations. The spectral range analyzed by these authors contained three lines of similar strength as the weak Fe~I~539.52~nm line. In general, the scatter in all lines is of the same magnitude, which suggests an instrumental origin (e.g., in determining the exact continuum level, or slightly variable spectral stray light, or slightly variable vignetting with time). The similar trends visible in EW of all these lines in the period from 1978 to 1992 support this conclusion, especially since these trends disappear after the larger grating was installed. However, we cannot exclude that neglecting 3D effects and the impact of the unresolved granular motion on the line profiles introduces an error to our models. This, together with the variability in stronger lines, should be studied with the next generation of irradiance models that will take this into account.

The only line that is insensitive to the velocity field and thus perfect for this type of modeling is Mn~I line, whose reconstructed time series of CD and EW agree quite well with the measured parameters. Reproducing the large solar cycle variation of the Mn I line with the SATIRE-S model without optimizing it in any way \citep[the single free parameter of the model remains
fixed to the value that was independently determined by][]{Ball:etal:2012} is a
success for the model and strengthens the assumption underlying it, namely
that variations in TSI and SSI (in the optical wavelength range) are caused
by the evolution of the solar surface magnetic field. Furthermore, the fact that SATIRE-S reproduces the correct level of the variation in the Mn line provides
strong support to the overall spectral profile of the  irradiance variability computed with SATIRE-S. On the one hand, \cite{Kok2015} have recently demonstrated that the
magnitude of the changes in the UV returned by SATIRE-S agrees well with the
most stable satellite measurements. On the other hand, since spectral lines determine the amplitude
and even the phase of the visible irradiance variations \citep{Shapiro:etal:2015}, our finding here indicates that the magnitude of the variability in
the visible is also adequate. Finally, the model also reproduces the changes in the TSI that are dominated by changes in the UV and the visible and are measured far more reliably than
changes in the spectral irradiance \citep{Kok:etal:2014}. Thus the model has now been independently tested in the three domains, which leaves little freedom for a significantly different profile of spectral irradiance variability. This further supports the conclusion of  \cite{Kok2014b} that the
SATIRE-S model currently presents the most realistic estimate of the solar
spectral irradiance variations and is to be preferred for climate studies.

We found a high correlation and a good agreement of the magnitude
of solar cycle variations between the observed and reconstructed
change in the Mn\,{\sc i} line parameters. This implies that the solar
cycle variations of the line \textit{can} be modeled by only
taking
into account changes in the surface distribution of the solar
magnetic features \citep{da:vi:2005,Vitas:etal:2009}. Since the solar disk coverage by faculae
increases from the minimum to the maximum of the solar cycle, the
disk-integrated line, in the 'Sun-as-a-star' spectrum, becomes
weaker during the solar cycle maximum. This explains why this
manganese line mimics the behavior of Ca\,{\sc ii}~K and Mg\,{\sc ii}~k lines,
which are well-known plage and faculae indicators. The influence of
sunspots is negligible. No additional temperature change in the
quiet-Sun component is necessary. 

\begin{acknowledgements}

We thank Rob Rutten for revising the manuscript very carefully and giving helpful instructions. This work was partly supported  by the German Federal Ministry of Education and Research under project 01LG1209A and by the BK21 plus program through the National Research Foundation (NRF) funded by the Ministry of Education of Korea.


\end{acknowledgements}

\bibliographystyle{aa}

\end{document}